\documentclass[aps,pre,twocolumn,showpacs]{revtex4-2}
\usepackage{amsmath}
\usepackage{amssymb}
\usepackage{epsf}
\usepackage{graphicx}
\usepackage{color}
\begin{document}
\title{Evidence for Additional Third-Order Transitions\\ in 
the Two-Dimensional Ising Model}
\author{Kedkanok Sitarachu}
\email[E-mail: ]{Kedkanok.Sitarachu@uga.edu}
\author{Michael Bachmann}
\email[E-mail: ]{bachmann@smsyslab.org}
\homepage[\\ Homepage: ]{http://www.smsyslab.org}
\affiliation{Soft Matter Systems Research Group, Center for
Simulational Physics, Department of Physics and Astronomy, University of 
Georgia, Athens, GA 30602, USA}
%
%\date{\today}
%
\begin{abstract}
We employ the microcanonical inflection-point analysis method, developed for 
the systematic identification and classification of phase 
transitions in systems of any size, to study the 
two-dimensional Ising model at various lattice sizes and in the thermodynamic 
limit. Exact results for the density of states, which were obtained by exact
algorithmic computation, provide evidence for higher-order transitions in 
addition to the well-studied second-order ferromagnetic-paramagnetic phase 
transition. An independent third-order phase transition is identified in the 
ferromagnetic phase, whereas another third-order transition resides in the 
paramagnetic phase. The latter is a dependent transition, i.e., it is 
inevitably associated with the critical transition, but it remains separate 
from the critical point in the thermodynamic limit. For a deeper insight into 
the nature of these additional transitions, a detailed analysis of spin 
clusters  
is performed.
\end{abstract}
%\pacs{05.70.Fh, 64.60.De, 64.70.-p, 82.35.Lr, 83.10.Tv}
\maketitle 
\section{Introduction}
The (Lenz-)Ising model was introduced about a century ago for studies of the 
impacts of attractive local spin-spin interaction upon macroscopic 
cooperative ordering across the entire system~\cite{lenz1,ising1}. As it 
turned out, the one-dimensional spin chain does not exhibit signs of a 
thermodynamic phase transition. It took almost two decades to solve the 
two-dimensional problem and to reveal the prominent second-order phase 
transition that separates the paramagnetic and the ferromagnetic 
phase~\cite{onsager1,kaufman1}. In the following decades, the simplicity and 
versatility of the model, an increased interest in understanding the origins 
of phase transitions, and the ever growing available computer power made the 
Ising model one of the most widely employed generic models for studies of 
complexity.  

Traditional theory dictates that phase transitions can only occur in the 
thermodynamic limit, which is where energetic and configurational response 
parameters tend to exhibit nonanalyticities at the transition point. From a 
modern point of view, this strict definition was mostly a reference to the 
mathematical tractability of complex problems. For the same reason, most 
studies of phase transitions were performed by employing canonical 
statistical analysis techniques. However, this approach is known to lead to 
problems in interpreting signals in response functions for systems of finite 
size. With fields like nano- and biosciences moving into 
the focus of statistical analysis, where cooperative system behavior is  
governed or at least strongly influenced by finite-size effects, the theory of 
phase transitions has to be extended and statistical 
analysis techniques appropriately adapted.

The significant evolution of computational resources throughout the last 
decades now allows algorithmic access to problems where a mathematical 
approach is not manageable. As desirable a rigorous treatment is, 
computational methods offer additional options for estimating or calculating 
quantities that are virtually inaccessible mathematically. One of the most 
interesting such quantity is the number (or density) $g(E)$ of microstates 
with system energy $E$. The logarithm of the density of 
states is commonly interpreted as the microcanonical entropy~\cite{gross1}
\begin{equation}
\label{eq:entropy}
S(E)=k_\mathrm{B}\ln g(E).
\end{equation}
The generalized microcanonical inflection-point 
analysis method was introduced for the study of systems of any 
size~\cite{qb1}. It rests on the 
principle of minimal sensitivity~\cite{stevenson1,stevenson2} in the
interplay 
between the configurational entropy $S(E)$ and the system 
energy $E$. In the microcanonical theory of phase transitions, these are 
considered the central quantities that govern effects competing 
with each other for dominance in the respective phases~\cite{gross1,mb1}. In 
consequence, their 
balance ensures a stable equilibrium state. In our method, the 
entropy and its derivatives with 
respect to energy are systematically analyzed to identify and 
classify transition signals uniquely~\cite{qb1}. 
The idea is similar to Ehrenfest’s 
approach to identifying and classifying phase transitions by means of 
nonanalyticities in derivatives of thermodynamic 
potentials~\cite{ehrenfest1}. However, the Ehrenfest scheme cannot be 
systematically 
extended to accommodate finite systems as nonanalyticities can only occur in 
the (hypothetical) thermodynamic limit.

We recently 
employed our method to analyze the phase behavior of various Ising 
systems~\cite{qb1,sb1,szb2}. As expected, the inflection-point analysis did 
not reveal transition signatures for the one-dimensional Ising chain. 
However, Ising strips and the two-dimensional (2D) Ising model on the square 
lattice exhibit a variety of transition signals. 
Particularly interesting are the higher-order transitions we found for the 2D 
Ising model in addition to the well-studied critical transition. According to 
our classification scheme, the critical transition is a 
second-order \emph{independent} transition, whereas an additional 
\emph{dependent} third-order transition was identified in the 
paramagnetic phase that is inevitably linked to the critical transition. 
It can be interpreted as the precursor of the critical transition in the 
disordered phase. Another independent transition is located in the ordered 
phase. In this paper, we provide evidence that these two additional 
transitions remain separate from the critical transition in the 
thermodynamic 
limit and thus can be considered phase transitions in the more general 
context provided by the microcanonical theory. By performing a detailed 
analysis of spin clusters, we also shed 
light on the character of these additional transitions.

It is worth noting that the microcanonical inflection-point analysis method 
has not only been 
successfully employed for spin systems, but 
also in studies of macromolecular systems~\cite{qb1,ab1,rizzi1}. It has 
proven useful as a foundation for a
better understanding of general geometric properties of phase 
transitions~\cite{pettini1,franzosi1} as well.

The paper is organized as follows: The Ising model, computational techniques, 
and the microcanonical analysis 
method are briefly reviewed in Section~\ref{sec:modmeth}.    
Results obtained by microcanonical inflection-point analysis are presented 
in Section~\ref{sec:micro}. Properties of the additional transitions 
identified by means of spin-cluster analyses are discussed in 
Section~\ref{sec:cluster}. The summary of the major results in 
Section~\ref{sec:sum} concludes the paper.
\section{Microcanonical statistical analysis and cluster simulations of the 
2D Ising model}
\label{sec:modmeth}
In the following, we briefly review the Ising model, 
the microcanonical inflection-point analysis method, and the simulation 
methodology used for the cluster analysis.
\subsection{Ising model}
In the two-dimensional Ising model~\cite{lenz1,ising1} with periodic boundary 
conditions and  
absent external magnetic field, 
the energy of the spin configuration $\mathbf{X}=(s_1,s_2,\ldots, s_N)$ 
with $N=L\times L$  
spins on a square lattice with edge lengths $L$ can simply be written as
\begin{equation}
\label{eq:ising}
E(\mathbf{X})=-J\sum\limits_{\langle i,j\rangle}s_is_j.
\end{equation}
Possible values of the spin orientation are $s_{i,j} = \pm 1$. The symbol
$\langle i,j\rangle$ indicates that only 
interactions of 
the  spins $s_i$ and $s_j$ are considered, if they are nearest neighbors on the 
lattice. The energy scale is fixed by the positive-valued coupling constant 
$J>0$ (ferromagnetic 
coupling). 
\subsection{Microcanonical inflection-point analysis method}
The microcanonical inflection-point analysis method, which utilizes 
the principle of least sensitivity~\cite{stevenson1,stevenson2}, was 
introduced to systematically identify and classify transition signals 
in systems of any size~\cite{qb1}. Like in canonical statistical analysis, 
the 
general assumption is that the interplay of entropy and energy governs the 
transition behavior. 

In our method, least-sensitive inflection points of 
$S(E)$, as defined in Eq.~(\ref{eq:entropy}), and its derivatives are used to 
identify phase transitions. We denote 
the derivatives as follows: $\beta(E)=dS(E)/dE$, $\gamma(E)=d^2S(E)/dE^2$, 
and $\delta(E)=d^3S(E)/dE^3$. Derivatives of higher order were not considered 
in this study.

As it turns out, it is useful to distinguish two types of transitions.
\emph{Independent} transitions are analogs of the conventional 
transitions and their occurrence does not depend on other cooperative 
processes in the system. This is in contrast to  
\emph{dependent} 
transitions, which are inevitably associated with an independent 
transition. These transitions only occur at a higher energy (i.e., usually in 
the less-ordered phase), and they are of higher 
order than the corresponding independent transition. Therefore, dependent 
transitions can be considered precursors of a major independent transitions. 
This may have noticeable consequences for applications: If a system 
currently 
in the disordered phase is adiabatically cooled down and a 
dependent transition signal is detected, a major phase transition is 
imminent upon further cooling.
\begin{figure*}
\centering
%\vspace{8mm}
\includegraphics[width = 0.8\textwidth]{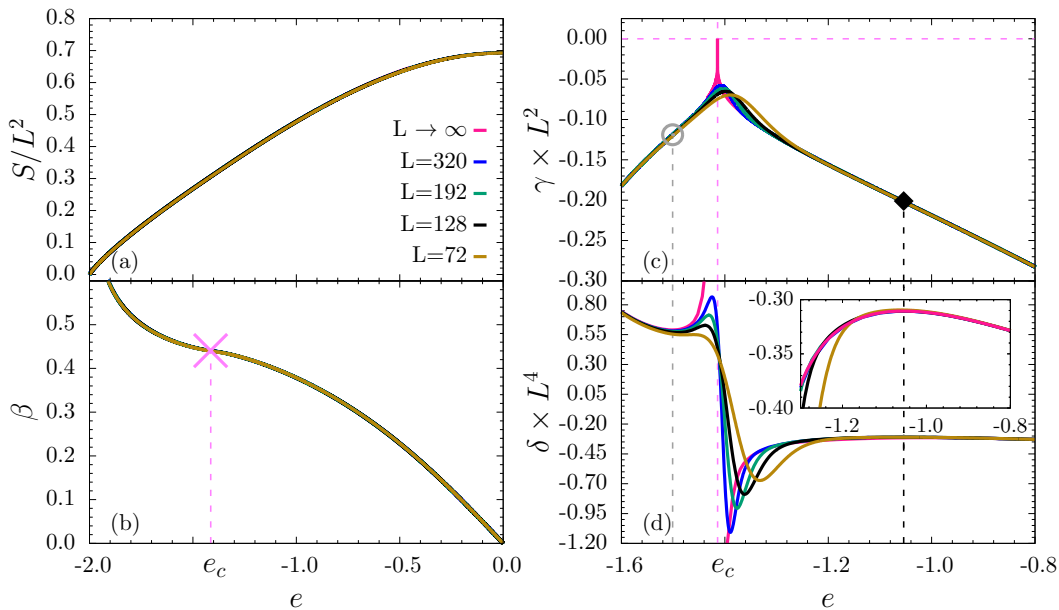}
\caption{\label{fig:Fig_Micro_curve} (a) Microcanonical entropy per 
spin $S(e)/L^2$ and its derivatives (b) $\beta(e)$, (c) $\gamma(e)L^2$, and 
(d) $\delta(e)L^4$ for various system sizes, plotted as functions 
of the energy per spin $e=E/L^2$. The dashed vertical lines 
mark the transition energies per spin associated with the three transitions 
found in the 2D Ising model. For reference, the critical energy per spin is
$e_\mathrm{c}\approx -1.414$. }
\end{figure*}

\emph{Independent} transitions are classified as of odd order $(2n-1)$, where 
$n$ is positive integer, if the inflection point at transition energy 
$E_\mathrm{tr}$ satisfies the condition
\begin{equation}
\left. \frac{d^{(2n-1)}S(E)}{dE^{(2n-1)}} 
\right|_{E=E_{\mathrm{tr}}}>0, 
\label{eq:odd_order_trans}
\end{equation}
whereas for even-order ($2n$) independent transitions
\begin{equation}
\left. \frac{d^{2n}S(E)}{dE^{2n}} \right|_{E=E_{\mathrm{tr}}}<0
\label{eq:even_order_trans}
\end{equation}
holds.
Inflection points are associated with even-order ($2n$) \emph{dependent} 
transitions, if
\begin{equation}    
\left. \frac{d^{2n}S(E)}{dE^{2n}} \right|_{E=E_\mathrm{tr}}>0, 
\label{eq:even_order_dep_trans}
\end{equation}
and odd-order $(2n+1)$ dependent transitions are characterized by
\begin{equation}
\left. \frac{d^{(2n+1)}S(E)}{dE^{(2n+1)}}
\right|_{E=E_\mathrm{tr}}<0. 
\label{eq:odd_order_dep_trans}
\end{equation}
For finite 2D Ising systems, it is convenient to use the exact algorithmic 
evaluation schemes introduced in Refs.~\cite{beale1,Haggkvist1} to determine 
the density of states. The latter method also allows 
for an extrapolation 
toward the thermodynamic limit, which will eventually 
permit us to decide whether or not transitions identified by means of the 
inflection-point method will survive in this limit. The 
derivatives of the microcanonical entropy are then obtained by numerical 
differentiation~\cite{szb2}.
\subsection{Wolff cluster algorithm}
For the study of cluster properties of the Ising system, we employed the 
Wolff 
single-cluster algorithm~\cite{Wolff}. 
Instead of performing single spin-flip Monte Carlo updates, in this method an 
entire cluster of spins is updated in a single step. This is most efficient 
near the critical point and in the subcritical ferromagnetic region, where 
the 
majority cluster coexists with smaller minority clusters.

In this simple yet powerful Monte Carlo method, one spin in the 
system is selected randomly.
Then, nearest-neighbor spins with the same orientation are identified and   
added to the (stochastic) Wolff cluster with probability
$p =1-\exp(-2\beta J)$,
where $\beta=1/k_\mathrm{B}T$ is the inverse thermal energy at temperature 
$T$ 
(the Boltzmann constant $k_\mathrm{B}$ was set to unity in the simulations 
and 
in the subsequent analysis). 
The process of adding spins to the Wolff cluster is repeated until all spins 
belonging to the same geometric cluster have been tested and the construction
of the Wolff cluster is complete. Eventually, all spins in this 
cluster are flipped.
For the identification of a geometric cluster, we used
the standard labeling technique introduced by Hoshen and 
Kopelman~\cite{Hoshen1}.
\section{Transition signals from microcanonical analysis}
\label{sec:micro}
Exact algorithmic methods~\cite{beale1,Haggkvist1} were employed to determine 
the densities of states of the 2D Ising model with 
periodic boundary conditions for lattice sizes with up to $320\times 320$ 
spins. The method by H\"aggkvist et al.~\cite{Haggkvist1} also allows to 
find the density of states in the thermodynamic limit ($L\to\infty$), 
which is key to judging whether or not the additional third-order transitions 
predicted previously~\cite{qb1,sb1,szb2} survive in this limit. 
Based on the exact data obtained from these algorithms, the microcanonical 
inflection-point analysis method was then used to identify 
transitions in the curves of the microcanonical entropy and its derivatives. 

The microcanonical results are shown in Fig.~\ref{fig:Fig_Micro_curve}. 
The quantities are properly rescaled to account for obvious system size 
dependence and plotted as functions of the energy per spin, $e=E/L^2$. 
Rescaled entropy and $\beta$ curves in Fig.~\ref{fig:Fig_Micro_curve}(a) 
and~\ref{fig:Fig_Micro_curve}(b), respectively, do not exhibit much system 
size dependence on the scales plotted. However, whereas there is no 
inflection point in the entropy, the $\beta$ curves do possess a unique 
least-sensitive inflection point, which indicates the well-studied critical 
transition separating the ferromagnetic from the paramagnetic phase. 
According to our
microcanonical classification scheme, it satisfies the criteria of an 
\emph{independent} second-order phase transition. In the thermodynamic limit, 
the critical transition energy per spin is $e_\mathrm{c}\approx -1.414$ and 
the critical temperature coincides with Onsager's result: 
$T_\mathrm{c}=2/\ln(1+\sqrt{2})\equiv 1/\beta(e_\mathrm{c})\approx 2.269$, 
as expected. Towards the thermodynamic limit ($L\to \infty$), the slope 
converges to zero, as can clearly be seen in the next derivative $\gamma(E)$, 
shown in Fig.~\ref{fig:Fig_Micro_curve}(c). Interestingly, the smooth peak 
visible for finite systems turns into a cusp in the thermodynamic limit. 
Consequently, the nondifferentiability of $\gamma$ at the critical transition 
energy leads to a discontinuity in the next-higher derivative $\delta$ 
[Fig.~\ref{fig:Fig_Micro_curve}(d)].

In addition to the critical transition, the microcanonical 
inflection-point analysis method identifies two additional transitions of 
higher order. An \emph{independent} third-order transition (fourth order for 
$L \leq 64$) is identified in the ferromagnetic phase. The corresponding 
least-sensitive inflection point in $\gamma$  
[Fig.~\ref{fig:Fig_Micro_curve}(c)] leads to a pronounced positive-valued 
local minimum in $\delta(e)$.
In the thermodynamic limit, 
the transition energy is $e_\mathrm{ind} \approx -1.502$, which 
corresponds to the transition temperature $T_\mathrm{ind} \approx 
2.229$.

Equally interesting is the occurrence of the \emph{dependent} third-order 
transition in the paramagnetic phase. As it is inevitably coupled to the 
critical transition, it can be imagined as a precursor of this major 
transition in the disordered phase. The least-sensitive inflection point in 
$\gamma(e)$, which converges to the transition energy 
$e_\mathrm{dep} = -1.053$ (corresponding to the transition 
temperature $T_\mathrm{dep} = 2.567$) in the thermodynamic limit, is 
characterized by a negative-valued peak in $\delta$. The inset in 
Fig.~\ref{fig:Fig_Micro_curve}(d) shows that this peak is also present in 
thermodynamic limit.

Whereas these additional transitions do not exhibit nonanalytic features in 
the way the critical transition does, the distinct signals indicating their 
existence survive in the thermodynamic limit and do not converge toward the 
critical point. This is a remarkable result as the subphases between them and 
the critical point create an ``atmosphere'' surrounding the critical 
transition. The dependent transition may potentially provide additional 
clues as to the loss of identity in the system when it approaches the 
critical point upon cooling. However, it is important to emphasize that the 
third-order transition in the ferromagnetic phase is independent of the 
critical transition and thus does not necessarily help understanding better 
the approach toward the critical transition upon adiabatic heating. It does 
not serve as a precursor of it in the way the dependent transition in the 
paramagnetic phase does.
\begin{figure}
\centering
\vspace{8mm}
\includegraphics[width = 0.47\textwidth]{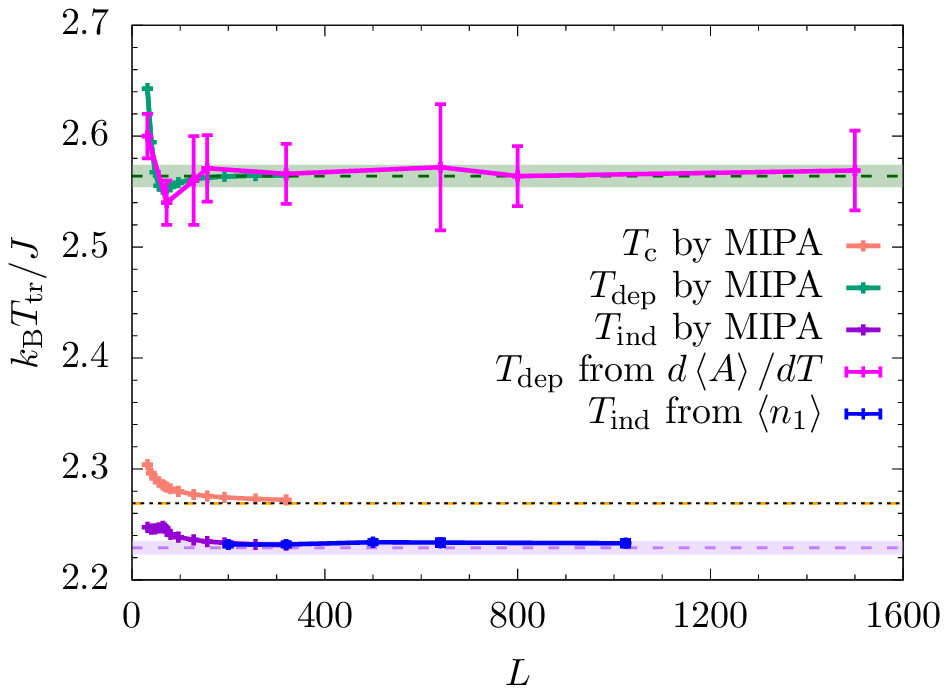}
\caption{\label{fig:Transition_B}
Transition temperatures $T_\mathrm{tr}$ obtained by microcanonical 
inflection-point analysis (MIPA) and cluster properties plotted 
as a function of $L$. Symbols mark the transition temperatures 
at finite system size (solid lines are only guides 
to the eye). Horizontal dashed lines are located at the transition 
temperatures in the thermodynamic limit ($L\to \infty$) found by 
microcanonical analysis. For reference, the 
critical temperature is $k_\mathrm{B}T_\mathrm{c}/J = 2/\ln(1+\sqrt{2}) 
\approx 2.269$. The small uncertainties in the microcanonical results 
originate from the numerical error in locating the transition signals due to 
the necessity of using discrete differences methods for calculating 
derivatives.}
\end{figure}

Figure~\ref{fig:Transition_B} contains the results for the transition 
temperatures obtained by microcanonical analysis for various lattice sizes 
and in the thermodynamic limit (dashed lines).  
It is important to note that the additional third-order transitions neither 
disappear nor converge toward the critical point in the thermodynamic limit. 
The transition temperatures remain well-separated from the critical 
temperature, but the microcanonical transition features do not develop 
into non-analyticities. Hence, these transitions are not phase transitions 
in the conventional Ehrenfest scheme. However, it should be reiterated that 
significant changes in system behavior in modern scientific problems and 
industrial applications~-- for many of which the thermodynamic limit is a 
nonsensical simplification~-- are not signaled by catastrophic changes in 
observables and data, but are rather subtle. Processes like 
folding and aggregation transitions of macromolecules, weather phenomena, 
swarm formation, and even synchronization in computer networks and social 
behavior occur on mesoscopic rather than macroscopic length scales. In fact, 
the early detection of sublying patterns leading to a catastrophic 
event may be more important and revealing than a thorough study of the major 
transition itself. 
\section{Analysis of spin clusters}
\label{sec:cluster}
We now discuss the results obtained by Wolff spin-cluster simulations and 
cluster analysis to shed more light on the system behavior associated with 
the 
additional transitions in the Ising model identified by microcanonical 
inflection-point analysis.
\subsection{Third-order dependent transition in the paramagnetic phase}
\begin{figure}
\centering
%\vspace{8mm}
\includegraphics[width = 0.45\textwidth]{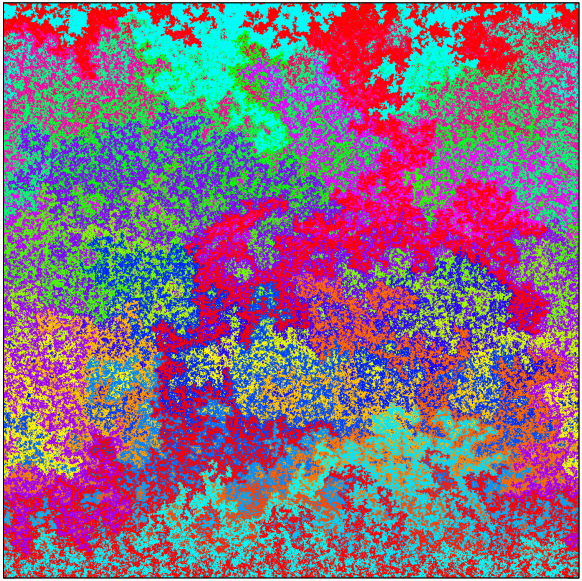}
\caption{\label{fig:config1500}
Clusters identified in a typical spin configuration on the 1500$\times$1500 
lattice in the paramagnetic phase at $T=2.605$, which is just above the 
dependent-transition point $T_\mathrm{dep} \approx 2.567$.
}
\end{figure}

In order to gain more insight into the nature of the additional third-order 
transitions identified for the 2D Ising model, cluster simulations were 
performed and cluster sizes analyzed by means of canonical statistical 
analyses of suitable order parameters. A typical example of a spin 
configuration on the square lattice with 1500$\times$1500 with all clusters 
colored differently is shown in Fig.~\ref{fig:config1500}.

The first quantity we take a closer look at is the average 
cluster size, $\langle A\rangle$, We define $A$ as the 
average size of clusters 
containing more than a single spin in a given spin configuration $\mathbf{X}$:
\begin{equation}
\label{eq:forA}
A=\frac{1}{n'}\sum\limits_{l'} C_{l'},
\end{equation}
where $l'$ labels the clusters with more than one spin, $C_{l'}$ is the 
number of spins in cluster $l'$, and 
$n'$ is the total number of clusters with more than one spin in 
$\mathbf{X}$. The statistical average is then obtained as
\begin{equation}
\label{eq:avA}
\langle A\rangle = \frac{1}{Z} \sum_{\mathbf{X}} 
A(\mathbf{X})e^{-E(\mathbf{X})/k_\mathrm{B}T},
\end{equation}
where $T$ is the canonical temperature and 
$Z=\sum_{\mathbf{X}}\exp[-E(\mathbf{X})/k_\mathrm{B}T]$ is the canonical 
partition function.

\begin{figure}
\centering
%\vspace{8mm}
\includegraphics[width = 0.47\textwidth]{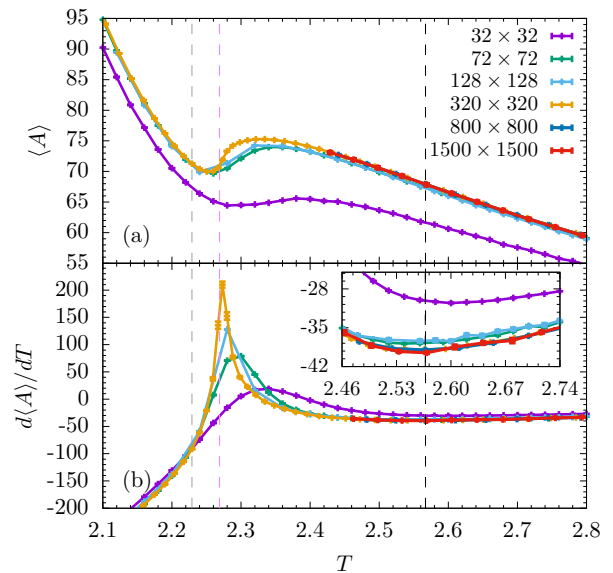}
\caption{\label{fig:Ave_all}(a) Average cluster size $\langle A\rangle$ and 
(b) derivative $d\langle A\rangle/dT$ as functions of temperature $T$
for the two-dimensional Ising model at various system sizes. The 
inset enlarges the area surrounding the dependent third-order transition in 
the paramagnetic phase. Note 
that cluster simulations for system sizes 800$\times$800 and 1500$\times$1500 
were only performed for temperatures $T\ge 2.46$.
The black dashed line indicates the location of the dependent third-order 
transition in the thermodynamic limit as obtained from microcanonical 
analysis, $T_\mathrm{dep} \approx 2.567$.
The other dashed lines locate the critical 
second-order and the subcritical independent third-order transition, 
respectively. }
\end{figure}

As mentioned, spin configurations at different temperatures were obtained in 
Wolff cluster 
simulations~\cite{Wolff}. At each temperature, up to $10^{8}$ spin 
configurations were generated. Spin clusters 
were labeled by means of the Hoshen-Kopelman method~\cite{Hoshen1} and the 
average cluster size $A$ in a given configuration 
was determined. The canonical average of this quantity 
and its derivative with respect to the temperature are plotted as functions 
of the temperature for various lattice sizes in Fig.~\ref{fig:Ave_all}.

Figure~\ref{fig:Ave_all}(a) shows that at low temperatures, in the 
ferromagnetic  
phase, the average cluster size decreases with increasing temperatures. 
Near the critical temperature, $\langle A \rangle$ exhibits 
a backbending pattern, which becomes more pronounced for larger lattices. 
For temperatures $T>T_\mathrm{c}$, i.e., in the paramagnetic phase, the 
average cluster size decreases again. The temperature derivative of the 
average cluster size, $d\langle A \rangle/dT$, is shown in 
Fig.~\ref{fig:Ave_all}(b).
It is a measure for the rate of change of the average cluster size with 
respect to the temperature. The curves for the different system sizes 
all show a prominent peak associated with the inflection point in the 
backbending pattern in Fig.~\ref{fig:Ave_all}(a). The peak location 
converges to the critical point, as expected. As a thermodynamic response 
quantity, it eventually becomes nonanalytic at the critical point in the 
thermodynamic limit. 

More interesting is the inflection point of $\langle A\rangle$ in the 
paramagnetic phase close to the dependent third-order transition identified 
by microcanonical analysis. It does not disappear even for the largest 
lattices simulated (1500$\times$1500). The curves of the derivative 
$d\langle A\rangle/dT$ exhibit a local minimum and there is no indication 
for it to flatten out in the thermodynamic limit. Its close proximity to the 
transition temperature of the dependent third-order transition 
$T_\mathrm{dep}\approx 2.567$ suggests that this feature is 
related to the transition. 

The decrease of the average cluster size with increasing temperature is 
expected in the paramagnetic phase. However, it is noteworthy that this 
decrease accelerates for temperatures $T<T_\mathrm{dep}$, 
before slowing down for $T>T_\mathrm{dep}$. This is 
an unexpected system behavior; the average cluster size could simply drop 
monotonously in the paramagnetic phase (in which case the third-order 
dependent transition would not exist). Although it seems to be a minor 
effect, this change of monotony is, in fact, an important signature of the 
catastrophic critical transition, because, as we have shown in the 
microcanonical analysis, these transitions are inevitably associated with 
each other. This means that the third-order dependent transition is a 
precursor of the critical transition in the paramagnetic phase, and~-- as our 
results from the cluster analysis show~-- is due to the change of the 
rate by which clusters decay in the disordered phase.

The estimates for the peak temperatures in $d\langle A\rangle/dT$ in the 
paramagnetic phase have already been included in Fig.~\ref{fig:Transition_B}. 
They clearly converge to the third-order dependent transition temperature
obtained by microcanonical analysis in the thermodynamic limit. Even for the 
finite lattices, the respective microcanonical estimate and the estimate from 
the cluster analysis are very close to each other, suggesting that the 
third-order transition signaled by microcanonical analysis is indeed due to 
the enhanced fluctuations about the average cluster size in this temperature 
region.
\subsection{Third-order independent transition in the ferromagnetic phase}
\begin{figure}
\centering
%\vspace{8mm}
\includegraphics[width = 0.47\textwidth]{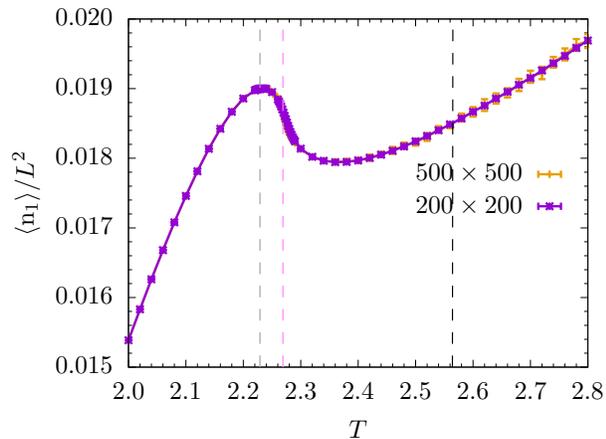}
\caption{\label{fig:onespin}
Average number of isolated spins per spin, $\langle n_1\rangle$, 
as a function of temperature for two lattice sizes. Vertical dashed lines are 
located at the 
transition temperatures of the 2D Ising model obtained by microcanonical 
analysis.
}
\end{figure}
\begin{figure}
\centering
%\vspace{8mm}
\includegraphics[width = 0.4\textwidth]{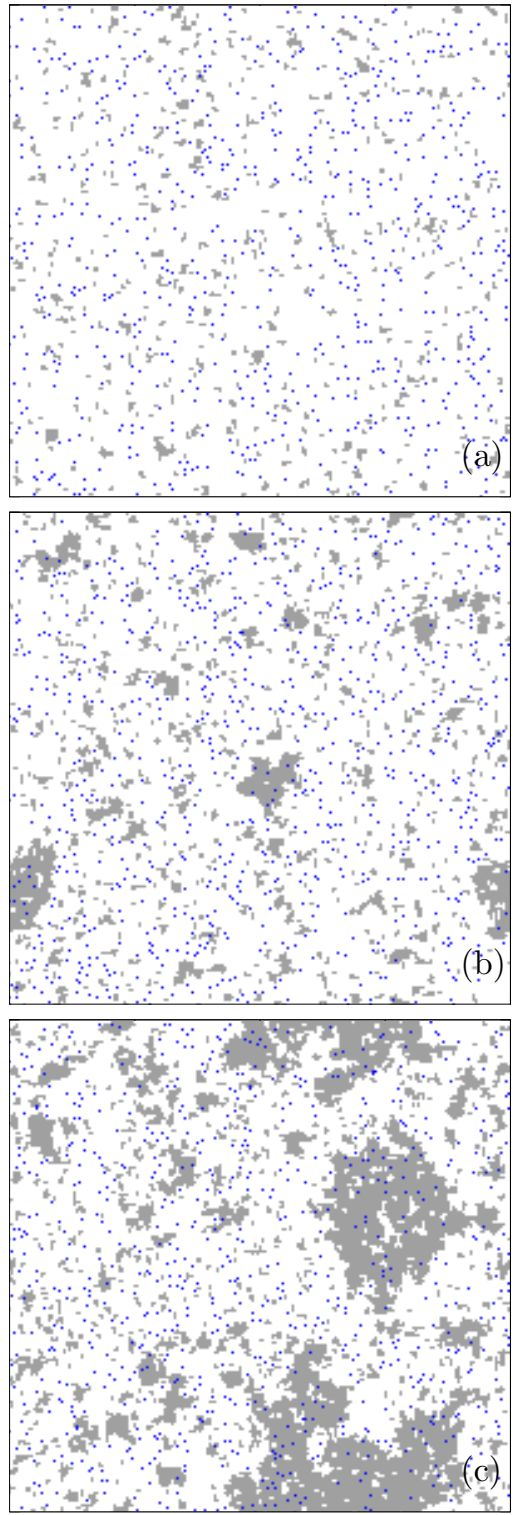}
\caption{\label{fig:config}
Representative Ising configurations on the 200$\times$200 lattice at (a) 
$T=2.10$, (b) $T=2.23$, and (c) $T=2.28$. In white areas, spins point up and 
in grey areas down. Isolated spins, independent of their orientation, are 
colored blue. The numbers of isolated spins divided by the total number 
of spins, $n_1$, are: (a) $0.0169$, (b) 
$0.0197$, and (c) $0.0179$.
}
\end{figure}
For the study of properties of the third-order independent transition in the 
ordered (ferromagnetic) phase, we look at signs of emerging disorder and 
entropic variability, which is dependent on the formation of minority 
clusters in this phase, where the ferromagnetic states are always dominated 
by a majority cluster. The simplest of these is obviously what we call the 
``single-spin cluster'', i.e., an isolated single spin surrounded by 
nearest-neighbor spins with opposite orientation. Figure~\ref{fig:onespin} 
shows plots of the statistical average of the number of isolated spins 
$\langle n_1\rangle$ as a function of temperature for two different lattice 
sizes. These results were also obtained in Wolff cluster simulations. Dashed 
vertical lines mark the transition points found by microcanonical 
analysis.

Most noteworthy is the peak near the third-order independent transition 
temperature $T_\mathrm{ind}\approx 2.229$ in the ferromagnetic phase and the 
subsequent drop toward the critical point. As expected, the number of 
isolated spins increases again with temperature in the paramagnetic phase.

The drop in the number of isolated spins in the ferromagnetic phase 
just below the critical point can be attributed to the dissolution 
of the majority cluster. Isolated spins serve as ``bond breakers.'' Their 
increased numbers and the subsequent recombination into clusters of 
smaller size with more and more rugged fractal boundaries occur near the 
third-order transition temperature. These cluster structures are not present 
in the pure ferromagnetic phase below $T_\mathrm{ind}$. In fact, clusters of 
intermediate size do not exist at all.

For example, an 
analysis of cluster sizes for the 500$\times$500 lattice revealed that 
near $T_\mathrm{ind}$ clusters of sizes in the range $10\%-70\%$ of the 
system 
size are completely absent. The population of intermediate-size cluster 
rapidly increases toward the critical temperature, though. The isolated spins 
help seed the formation of these clusters.
Representative configurations on the 200$\times$200 Ising lattice are shown 
in Fig.~\ref{fig:config} for temperatures (a) below $T_\mathrm{ind}$, (b) 
near $T_\mathrm{ind}$, and (c) close to $T_\mathrm{c}$. 
Note that the transition at 
$T_\mathrm{int}\approx 2.229$ is an independent transition, i.e., it is not 
associated with the critical transition.

In Fig.~\ref{fig:Transition_B}, we have already included the transition 
temperatures of this transition for various lattice sizes. Even for the 
largest system simulated in this phase, 1024$\times$1024 spins, the peak 
temperature read off 
from $\langle n_1\rangle$ is very close to the microcanonical estimate 
at this system size. Most importantly, the transition temperature estimates 
for 1024$\times$1024 and even smaller lattice sizes, are located  well within 
the narrow uncertainty region of the microcanonical estimate for the 
transition temperature in the thermodynamic limit. This increases the 
confidence that the peaking in the average 
number of isolated spins in the ferromagnetic phase is a major feature of the 
system behavior in the vicinity of this third-order transition. As 
expected, the finite-size transition temperatures of this additional 
transition do not converge toward the critical 
temperature, but remain separate from the critical point even in the 
thermodynamic limit.
\section{Summary}
\label{sec:sum}
The purpose of this study was twofold: First, it was necessary to verify that 
the recently found additional phase transitions in the two-dimensional Ising 
model flanking the critical transition remain present and well-separate from 
the critical transition in the thermodynamic limit. This goal could be 
achieved by microcanonical inflection-point analysis of the microcanonical 
entropy and the relevant derivatives~\cite{qb1} in the thermodynamic limit. 
This was made possible by means of the exact enumeration method for the 
density of states of the Ising model introduced by H\"aggkvist et 
al.~\cite{Haggkvist1}. 

The second objective was to find evidence of the 
third-order transitions in the way the Ising system forms clusters in both 
the paramagnetic and the ferromagnetic phase. For this purpose, extensive 
Wolff single-cluster simulations~\cite{Wolff} for lattice systems with up to 
1500$\times$1500 spins were performed and suitably introduced order 
parameters measured.

It turned out that the 
fluctuations of the average cluster size (excluding isolated single spins) 
become extremal at about the temperature of the third-order dependent 
transition in the paramagnetic phase. This suggests that a 
collective pre-ordering of spins occurs in this temperature region in the 
disordered phase as a precursor of the critical transition. 

In the ferromagnetic phase, the average number of isolated spins peaks at the 
independent third-order transition temperature that was identified by 
microcanonical analysis. Here, the increased number of such ``seeds'' of 
disorder in the ferromagnetic phase enables the formation of critical 
clusters once the critical point is approached. 

These results are encouraging and may initiate a search for higher-order 
transitions in other systems as well. Our analysis also shows that the study 
of sublying transitions in ordered and disordered phases can lead to a better 
understanding of the system-inherent reasons leading to major phase 
transitions. Dependent transitions, which are inevitably coupled to a major 
transition, are precursors of imminent global ordering processes in the 
disordered phase. The understanding of such precursor transitions may aid 
predicting significant ordering effects such as cooperativity and 
synchronization in complex systems before they actually happen. 
\begin{acknowledgments}
We thank the Georgia Advanced Computing Resource Center (GACRC) at the 
University of Georgia for providing computational resources.
\end{acknowledgments}

\end{document}